\documentclass[twocolumn,showpacs,preprintnumbers,amsmath,amssymb,prb,superscriptaddress]{revtex4-1}

\usepackage{amssymb}
\usepackage{graphicx}
\usepackage{bm}
\usepackage{natbib}

\begin{document}
\title{Distinct magnetic spectra in the hidden order and antiferromagnetic phases in URu$_{2-x}$Fe$_x$Si$_2$}

\author{Nicholas P. Butch}
\email{nicholas.butch@nist.gov}
\affiliation{NIST Center for Neutron Research, National Institute of Standards and Technology, 100 Bureau Drive, Gaithersburg, MD 20899}
\affiliation{Center for Nanophysics and Advanced Materials, Department of Physics, University of Maryland, College Park, MD 20742}
\author{Sheng Ran}
\affiliation{Department of Physics, University of California - San Diego, 9500 Gilman Drive, La Jolla, CA 92093}
\affiliation{Center for Advanced Nanoscience, University of Calfornia – San Diego, 9500 Gilman Drive, La Jolla, CA 92093}
\author{Inho Jeon}
\affiliation{Center for Advanced Nanoscience, University of Calfornia – San Diego, 9500 Gilman Drive, La Jolla, CA 92093}
\affiliation{Materials Science and Engineering Program, University of Calfornia – San Diego, 9500 Gilman Drive, La Jolla, CA 92093}
\author{Noravee Kanchanavatee}
\altaffiliation{Present address: Department of Physics, Chulalongkorn University, Pathumwan, Thailand}
\affiliation{Department of Physics, University of California - San Diego, 9500 Gilman Drive, La Jolla, CA 92093}
\affiliation{Center for Advanced Nanoscience, University of Calfornia – San Diego, 9500 Gilman Drive, La Jolla, CA 92093}
\author{Kevin Huang}
\altaffiliation{Present address: State Key Laboratory of Surface Physics, Department of Physics, Fudan University, Shanghai, China}
\affiliation{Center for Advanced Nanoscience, University of Calfornia – San Diego, 9500 Gilman Drive, La Jolla, CA 92093}
\affiliation{Materials Science and Engineering Program, University of Calfornia – San Diego, 9500 Gilman Drive, La Jolla, CA 92093}
\author{Alexander Breindel}
\affiliation{Department of Physics, University of California - San Diego, 9500 Gilman Drive, La Jolla, CA 92093}
\affiliation{Center for Advanced Nanoscience, University of Calfornia – San Diego, 9500 Gilman Drive, La Jolla, CA 92093}
\author{M. Brian Maple}
\affiliation{Department of Physics, University of California - San Diego, 9500 Gilman Drive, La Jolla, CA 92093}
\affiliation{Center for Advanced Nanoscience, University of Calfornia – San Diego, 9500 Gilman Drive, La Jolla, CA 92093}
\affiliation{Materials Science and Engineering Program, University of Calfornia – San Diego, 9500 Gilman Drive, La Jolla, CA 92093}
\author{Ryan L. Stillwell}
\affiliation{Lawrence Livermore National Laboratory, 7000 East Ave., Livermore, CA 94550}
\author{Yang Zhao}
\affiliation{NIST Center for Neutron Research, National Institute of Standards and Technology, 100 Bureau Drive, Gaithersburg, MD 20899}
\affiliation{Department of Materials Science and Engineering, University of Maryland, College Park, Maryland 20742, USA}
\author{Leland Harriger}
\author{Jeffrey W. Lynn}
\affiliation{NIST Center for Neutron Research, National Institute of Standards and Technology, 100 Bureau Drive, Gaithersburg, MD 20899}
\date{\today}

\begin{abstract}
We use neutron scattering to compare the magnetic excitations in the hidden order (HO) and antiferromagnetic (AFM) phases in URu$_{2-x}$Fe$_{x}$Si$_2$ as a function of Fe concentration. The magnetic excitation spectra change significantly between $x=0.05$ and $x=0.10$, following the enhancement of the AFM ordered moment, in good analogy to the behavior of the parent compound under applied pressure. Prominent lattice-commensurate low-energy excitations characteristic of the HO phase vanish in the AFM phase. The magnetic scattering is dominated by strong excitations along the Brillouin zone edges, underscoring the important role of electron hybridization to both HO and AFM phases, and the similarity of the underlying electronic structure. The stability of the AFM phase is correlated with enhanced local-itinerant electron hybridization.

\end{abstract}

\maketitle

The nature of the Hidden Order (HO) phase of URu$_2$Si$_2$ is a longstanding challenge for condensed matter physics. The phase transition is characterized by a large entropy change at 17.5~K along with features in electrical resistivity and magnetic susceptibility \cite{Schlabitz86,Palstra85,Maple86}. Along with multiple studies that show a gap opening in the charge \cite{Park12,Aynajian10,Schmidt10,Lobo15} and spin \cite{Broholm87,Bourdarot10,Wiebe07} excitation spectra, these properties indicate that the HO phase involves a rearrangement of the high-temperature correlated electronic state composed of interacting itinerant and localized \emph{f}-electrons. Nonetheless, identification of the static order parameter in the HO phase remains elusive \cite{Mydosh11}, and even a proper description of the \emph{f}-electron state on uranium is controversial \cite{Jeffries10,Wray15}.

\begin{figure}
\begin{center}
\includegraphics[width=3.4in]{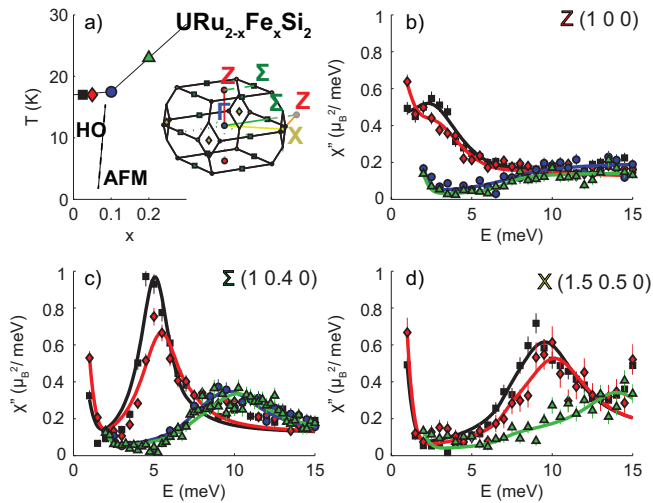}
\end{center}
\caption{Composition dependence of the ground state magnetic excitations in the HO and AFM phases. a) The phase diagram shows the transition temperatures for four different Fe concentrations: $x = 0.025$ (black squares), 0.05 (red diamonds), 0.1 (blue circles), and 0.2 (green triangles). The reciprocal space map highlights important directions. Strong low-energy magnetic fluctuations are suppressed in the AFM phase, as seen at b) the \textbf{Z} point, or antiferromagnetic zone center; c) the $\pmb{\Sigma}$ point on the zone edge, where some spectral weight has shifted to higher energy; and d) the \textbf{X} point on the diagonal edge. The inelastic spectra for $x=0.1$ and 0.2 nearly overlap. Shapes and colors of data points correspond to $x$ values from a). Lines represent fits described in the text. Data were collected on BT-7.}
\label{xcom}
\end{figure}

The unusual spin excitation spectrum of URu$_2$Si$_2$ offers some clues to the underlying interactions in both the correlated paramagnetic and HO phases. The most prominent, and most studied, features are excitations at Brillouin zone (BZ) face center \textbf{Z} and at an incommensurate point $\pmb{\Sigma}$, which sits on a BZ edge \cite{Broholm87,Bourdarot10,Wiebe07} (see Fig.~\ref{xcom}a). The latter resonance actually represents part of a ring of excitations that approximately follows the BZ edge, which can be attributed to interband transitions of the correlated electron bands \cite{Butch15}. Meanwhile, the presence of a sharp magnetic dispersion with a mininum at \textbf{Z} has been used as evidence that the HO phase breaks spatial symmetry in a manner similar to antiferromagnetic (AFM) order\cite{Buhot14,Kung15}, despite the lack of a component of static dipolar magnetic order \cite{Amitsuka07,Das13,Ross14}. True AFM order is stabilized by applied pressure \cite{Amitsuka99,Butch10}, and measurements at the \textbf{Z} and $\pmb{\Sigma}$ points show that the magnetic excitations energies change discontinuously in the AFM phase \cite{Villaume08,Bourdarot14} - in particular, the \textbf{Z} excitation gap appears to close. However, the momentum dependence of these excitations remains unknown.

Fortunately, Fe substitution appears to mimic the effects of applied pressure by stabilizing AFM order \cite{Kanchanavatee11}. Neutron diffraction measurements suggest that the phase diagrams are analogous at low Fe concentration and low applied pressure, albeit with a larger moment in the Fe case \cite{Das15}. Thermal expansion measurements on URu$_{1.9}$Fe$_{0.1}$Si$_2$ demonstrate that the HO-AFM phase boundary passes through this concentration
\cite{Ran16}, similar to what was seen in URu$_2$Si$_2$ under pressure \cite{Villaume08}. Optical conductivity \cite{Hall15} and muon spin rotation \cite{Wilson16} measurements also support the existence of an AFM phase, but there are some disagreements about the exact location of the HO-AFM phase boundary.

To probe the momentum dependence of the magnetic excitation spectrum in the AFM phase, we performed inelastic neutron scattering measurements on single crystals of URu$_{2-x}$Fe$_{x}$Si$_2$ in both the HO and AFM phases. We find that Fe substitution broadens and slightly suppresses the dispersive excitations at \textbf{Z}, and that they vanish in the AFM phase, where they are replaced by a 5~meV gap. The phase boundary separating HO from AFM is located between $x=0.05$ and $x=0.10$. The lack of conventional spin waves in the AFM phase calls into question the use of \textbf{Z}-point excitations as proof of zone folding in the HO phase. In contrast, the incommensurate excitations at $\pmb{\Sigma}$ and around the zone edge remain, although the energy gap increases to 8~meV. The same hybridized interband correlations serve as a background to both HO and AFM phases, and both order parameters gap the spin fluctuation spectrum similarly, suggesting similar electronic structures. The AFM phase is stabilized by increased electron hybridization.

Single crystals of URu$_{2-x}$Fe$_{x}$Si$_2$ that were synthesized via the Czochralski technique in a continuously-gettered tetra-arc furnace. Neutron scattering measurements were carried out at the NIST Center for Neutron Research on the BT-7 thermal triple axis spectrometer \cite{Lynn12} with 14.7~meV final energy, the NG-5 cold triple axis spectrometer with 3.7~meV final energy, and the NG-4 disc-chopper spectrometer \cite{Copley03}. Data analysis was performed using the DAVE software suite \cite{Azuah09}. Throughout this paper, error bars associated with measurements and fits correspond to one standard deviation.

The body-centered tetragonal (BCT) lattice of URu$_2$Si$_2$ has a BZ with high-symmetry points of importance to the magnetic excitations: BZ center $\pmb{\Gamma}$, horizontal face center \textbf{X}, vertical face center \textbf{Z} and horizontal edge center $\pmb{\Sigma}$.  The \textbf{Z} point also represents the AFM zone center, and becomes equivalent to $\pmb{\Gamma}$ when the lattice symmetry is reduced to simple tetragonal in the AFM phase, and perhaps in the HO phase. Neutron diffraction shows that the Fe-stabilized AFM magnetic structure is equivalent to that in URu$_2$Si$_2$ under pressure \cite{Amitsuka99}.

The effects of Fe substitution on the magnetic excitations at the \textbf{Z}, $\pmb{\Sigma}$, and \textbf{X} points clearly delineate the HO and AFM phases. Figure~\ref{xcom}a shows the magnetic phase diagram of URu$_{2-x}$Fe$_{x}$Si$_2$ and the transition temperatures of the studied samples. The $x$-dependence of the magnetic excitations in the ground state is depicted via scans at constant momentum $Q$ and varying energy $E$ at the b) \textbf{Z}, c) $\pmb{\Sigma}$, and d) \textbf{X} points. The magnetic scattering intensity has been normalized to absolute units by comparison to phonon scattering and to the $x$-independent paramagnetic fluctuation intensity at $\pmb{\Sigma}$. The susceptibility derives from scaling by the population factor $1 - e^{-\frac{E}{k_BT}}$ that reduces the low-$E$ contribution at higher temperature $T$. The characteristic sharply peaked excitations of the HO phase are weakened in the AFM phase. Most striking is the \textbf{Z} point, where the prominent excitation is replaced by a gap in the magnetic fluctuations spectrum. The $E$-dependence of the inelastic scattering can be described by a step function with $x$-dependent inflection point, and a Lorentzian function centered at finite energy, which is present only in the HO phase. The abrupt change in the magnetic excitation spectrum between $x=0.05$ and $x=0.10$ places the phase boundary in between those concentrations. At $\pmb{\Sigma}$, the excitations weaken and increase in energy, although here a broad Lorentzian persists in the AFM phase. At \textbf{X}, the excitations vanish in the measured $E$ range, apparently shifting to higher $E$ outside the measurement window. No prominent low-$E$ excitations have been detected in cold neutron measurements, and a 0.15~$\mu_B^2/$~meV upper bound is estimated for any excitations below 1.5~meV. Neutron measurements were not sensitive to the two transitions in $x=0.1$ that were observed in thermal expansion \cite{Ran16}. The slight shifts in $E$ at low $x$ at $\pmb{\Sigma}$ and \textbf{Z} mimic the effects of applied pressure in URu$_2$Si$_2$ \cite{Bourdarot14}.

\begin{figure}
\begin{center}
\includegraphics[width=3.4in]{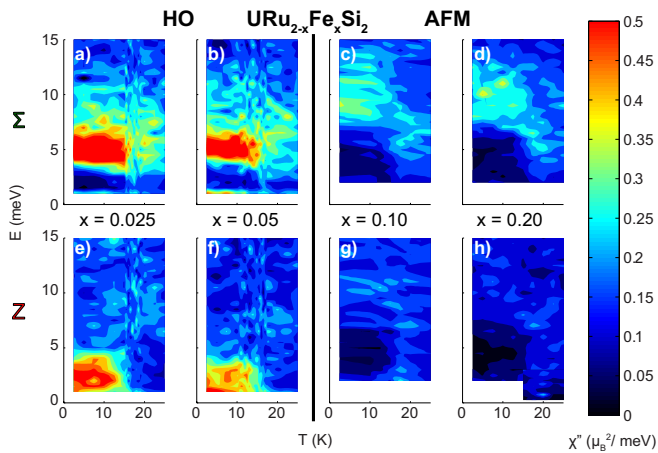}
\end{center}
\caption{Temperature dependence of the magnetic excitations in the a,b,e,f) HO and c,d,g,h) AFM phases. At both the zone edge $\pmb{\Sigma}$ (a-d) and AFM zone center \textbf{Z} (e-h), broad magnetic fluctuations shift to higher energy below the respective ordering temperatures. The opening of the gap is abrupt as a function of temperature in all cases. In the AFM phase in $x=0.2$, the magnetic excitation intensity at \textbf{Z} is reduced overall, but a temperature-dependent magnetic gap remains a telltale feature. Data were collected on BT-7, and the low-energy data in h on NG-5.}
\label{Tcom}
\end{figure}

The temperature dependence of the magnetic excitations follows the onset of the ordered phase as determined by neutron diffraction and thermodynamic measurements \cite{Das15}. This is shown in Fig.~\ref{Tcom}: in the HO phase, at both $\pmb{\Sigma}$ and \textbf{Z}, the excitation spectrum changes from weak, low-energy excitations in the correlated paramagnetic state to higher-energy excitations in the HO and AFM phases, with an energy gap comparable to that observed in URu$_2$Si$_2$. In the AFM phase, the intensity of the excitations is weaker and the peak energies are higher. At both $\pmb{\Sigma}$ and \textbf{Z}, the gap value can be roughly defined as the inflection point in the $E$-dependence of the intensity, with a value of 7-8~meV.  These values are consistent with the published pressure dependence in URu$_2$Si$_2$ \cite{Villaume08,Bourdarot14}, but these values are larger than those determined from optical conductivity measurements on Fe-substituted samples \cite{Hall15}.

\begin{figure}
\begin{center}
\includegraphics[width=3.4in]{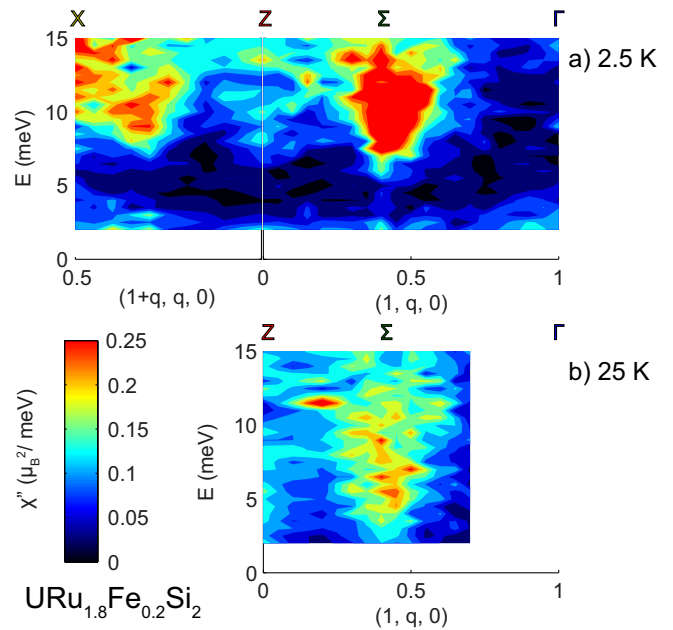}
\end{center}
\caption{Momentum dependence of the magnetic excitations in the AFM phase for $x =0.2$. a) In the ground state, along the (100) direction, from \textbf{Z} to $\pmb{\Gamma}$, the most prominent excitations are at the zone edge at $\pmb{\Sigma}$ where the gap is smallest, as is the case in the HO phase. Also along the (110) direction, from \textbf{Z} to \textbf{X}, the gap minimum and strongest magnetic intensity occur near the zone edge. b) Along the (100) direction in the paramagnetic phase, the energy gap decreases significantly, but the intensity is still maximum at $\pmb{\Sigma}$.  Data were collected on BT-7.}
\label{Qcom}
\end{figure}

\begin{figure}
\begin{center}
\includegraphics[width=3.4in]{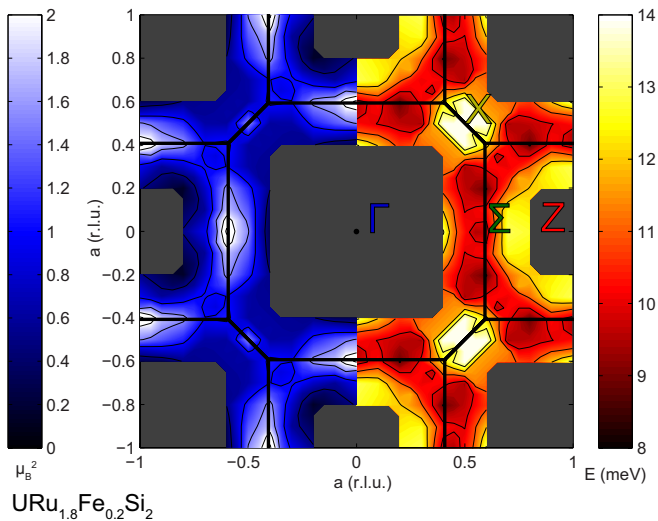}
\end{center}
\caption{Reciprocal space map of the excitations in the AFM phase: the energy-integrated magnetic intensity (left) and the magnetic dispersion (right). As in the HO phase, the maximum intensity and energy minima follow the zone boundary.}
\label{maps}
\end{figure}

The detailed momentum dependence of the magnetic excitations in the AFM phase is shown in Fig.~\ref{Qcom}. The oft-studied magnetic dispersions along the (100) direction and its symmetry equivalents, are presented in Fig.~\ref{Qcom}a. The most prominent excitations are centered at the BCT zone edge $\pmb{\Sigma}$, which coincides with the dispersion minimum. As in URu$_2$Si$_2$, these excitations disperse steeply upward toward both \textbf{Z} and $\pmb{\Gamma}$. Any dispersion centered on \textbf{Z} is difficult to conclusively define due to the weakness of the excitations. The \textbf{Z}-\textbf{X} direction is qualitatively similar, with a dispersion minimum at $q \approx 0.35$, as in the parent compound. Thus it is apparent that the AFM and HO phases both have in common a ring of strong magnetic excitations that traces the BCT zone edge. The high-temperature fluctuations of this ring are also analogous, as shown in Fig.~\ref{Qcom}b. At 25~K in the correlated paramagnetic phase, the excitations are broad and at lower $E$ but remain centered on \textbf{Z} and $\pmb{\Sigma}$. The momentum dependence of the ring in the AFM phase is remarkably similar to that in the HO phase in URu$_2$Si$_2$, as summarized in Fig.~\ref{maps}, emphasizing that the excitations traverse the zone edge in the same manner as in the parent compound \cite{Butch15}, although the $E$-integrated fluctuating moment is reduced by a factor of $\sqrt{2}$.

These results underscore the interpretation of the incommensurate excitations as those due to interband scattering in URu$_2$Si$_2$. This description is consistent with the $x$ dependence of the specific heat, magnetization, and electrical resistivity, which remain similar in this range of Fe concentration \cite{Kanchanavatee11}, meaning that the correlated electron state and the details of its band structure do not change dramatically. Within the context of an interband hopping model \cite{Butch15}, the implication is that a small electron pocket remains at $\pmb{\Gamma}$ and a large hole pocket remains centered on \textbf{Z}. The change in resonance energy signifies a 4~meV increase in the indirect hybridization gap that appears to be tied to the change in ground state. Because the momentum dependence of the interband transitions remains the same, without invoking changes in the uncorrelated band structure, the gap increase can be simply related to an increase in the local-itinerant hybridization potential, which may be naturally expected in the context of shrinking interatomic spacing due to chemical pressure \cite{Kanchanavatee11}. This enhanced hybridization may represent a crucial component of what determines the relative stability of the HO and AFM phases.

There are also other apparent changes in the band structure. The obvious lack of prominent, sharply dispersing excitations at \textbf{Z} distinguishes the AFM phase from HO. If the electronic structures in the Fe-tuned AFM and HO phases are similar as they are under pressure\cite{Hassinger10}, then the absence of prominent excitations cannot be due to the absence of the small hole-like band at \textbf{Z} in the AFM phase. If the weak excitations at \textbf{Z} also stem from interband transitions, then the hybridization gap at \textbf{Z} has also increased by 4~meV, meaning that the hybridization of different bands has been affected similarly. In the HO phase, magnetic excitations disperse near the \textbf{X} point, albeit weakly \cite{Butch15}. In the AFM phase, on the other hand, these excitations are not observed (Fig.~\ref{xcom}d). It is possible that they are also pushed to higher $E$, where broadening and the presence of phonons make identification difficult, but a compelling alternative is that these excitations vanish in the AFM phase, yielding another clue to the stability of the HO phase. Indeed, the \textbf{X} point was identified in angle-resolved photoemission spectroscopy (ARPES) studies as being near possible nesting or hotspot wavevectors \cite{Boariu13,Meng13}. It is therefore a priority to confirm the hybridization gap increase and the inferred band structure in URu$_{1.8}$Fe$_{0.2}$Si$_2$ via ARPES and quantum oscillations measurements.

Another important feature of the AFM magnetic excitations is the absence of strong low-$E$ spin waves near \textbf{Z}. This contrasts with the readily apparent \textbf{Z}-centered excitations in the HO phase, which have been interpreted as evidence that the HO phase breaks spatial symmetry in a manner similar to the AFM phase, most recently in the context of Raman scattering \cite{Buhot14,Kung15}. The absence of similar excitations in the AFM phase implies that the \textbf{Z}-centered magnetic excitations in the HO phase are not a sure signature of BCT symmetry-breaking in the HO phase. A more conventional interpretation of the data is that the \textbf{Z}-centered interband transitions are actually the AFM spin waves (Fig.~\ref{Tcom}d, Fig.~\ref{Qcom}a), but this is not straightforward either; it contradicts the trend in peak energy observed in the HO phase (Fig.~\ref{xcom}b) and under pressure \cite{Bourdarot14}, and yields an excitation energy scale that is several times greater than the transition temperature of 20~K, whereas the HO phase is more conventional in this regard. Either way, the AFM phase has unconventional character and is interesting in its own right.

It should be noted that even in the HO phase, the \textbf{Z}-centered excitations make up a negligible fraction of the total fluctuating moment when compared to the much stronger incommensurate excitations that occupy a much larger fraction of reciprocal space. Their presence or absence would affect the bulk physical properties of the HO and AFM phases only subtly, as experiment shows \cite{Jeffries07,Hassinger08}. This highlights that the AFM phase is in many ways just as subtle as the HO itself, and it is advantageous that the AFM order parameter is already known. Developing a proper theoretical understanding of the weak AFM excitations should be more tractable than identifying HO directly, and may provide an alternate route to understanding the complicated electron interactions at the root of both phenomena.

Our measurements strongly support the analogy between the Fe-substituted and pressure-tuned phase diagrams, extending the experimental possibilities for studying the AFM phase, which otherwise remains technically challenging in the parent compound due to the constraints inherent to pressure cells. In addition to measurements sensitive to band structure, it will be interesting to compare the AFM properties in high magnetic fields, and the detailed characteristics of the low-temperature superconductivity. The specifics of the AFM magnetic excitation spectrum will also provide useful constraints for theoretical descriptions of the AFM phase, as well as the HO phase that it borders.

Finally, we note that our inelastic spectra are consistent with the unpublished data of T. J. Willams and coworkers \cite{Williamsnote}. Also, a recent study concludes that in the pressure-induced AFM phase in URu$_2$Si$_2$, the inelastic excitations at \textbf{Z} remain at finite energy \cite{Williams16}, which differs significantly from the behavior reported in Refs.~\onlinecite{Villaume08,Bourdarot14}.

We thank M. Janoschek and J. S. Helton for helpful discussions, and T. J. Williams for sharing his unpublished neutron scattering data. Single crystal growth and characterization at UCSD were supported by the US Department of Energy, Office of Basic Energy Sciences, Division of Materials Sciences and Engineering, under Grant No. DEFG02-04-ER46105. Low temperature measurements at UCSD were sponsored by the National Science Foundation under Grant No. DMR 1206553.

\end{document}